\documentclass[sigconf,natbib=true,anonymous=false]{acmart}

\usepackage{multirow}
\usepackage{caption}
\usepackage{subcaption}
\usepackage{xspace}
\usepackage{xcolor}
\usepackage{enumitem}
\usepackage{array}

\newtheorem{definition}{Definition}
\newtheorem{observation}{Observation}

\newcommand{\paratitle}[1]{\smallskip\noindent\textbf{#1}}
\newcommand{\ie}{\emph{i.e.,}\xspace}
\newcommand{\eg}{\emph{e.g.,}\xspace}

\AtBeginDocument{%
  \providecommand\BibTeX{{%
    \normalfont B\kern-0.5em{\scshape i\kern-0.25em b}\kern-0.8em\TeX}}}

\setcopyright{acmcopyright}
\copyrightyear{2024}
\acmYear{2024}
\acmDOI{XXXXXXX.XXXXXXX}

  \acmConference[Conference acronym 'XX]{Make sure to enter the correct conference title from your rights confirmation emai}{June 03--05,
  2018}{Woodstock, NY}
\acmPrice{15.00}
\acmISBN{978-1-4503-XXXX-X/18/06}

\begin{document}



\title{Recommender for Its Purpose: Repeat and Exploration in Food Delivery Recommendations}


\author{Jiayu Li}
\email{jy-li20@mails.tsinghua.edu.cn}
\affiliation{
  \institution{ DCST, Tsinghua University}
  \state{Beijing}
  \country{China}
}

\author{Aixin Sun}
\authornote{Aixin Sun and Min Zhang are the corresponding authors.}
\email{axsun@ntu.edu.sg}
\affiliation{
  \institution{SCSE, Nanyang Technological University}
  \country{Singapore}
}

\author{Weizhi Ma}
\email{mawz@tsinghua.edu.cn}
\affiliation{
  \institution{ AIR, Tsinghua University}
  \state{Beijing}
  \country{China}
}

\author{Peijie Sun}
\email{sun.hfut@gmail.com}
\affiliation{
  \institution{ DCST, Tsinghua University}
  \state{Beijing}
  \country{China}
}

\author{Min Zhang}
\authornotemark[1]
\email{z-m@tsinghua.edu.cn}
\affiliation{
  \institution{ DCST, Tsinghua University}
  \state{Beijing}
  \country{China}
}
\begin{abstract}
Recommender systems have been widely used for various scenarios, such as e-commerce, news, and music, providing online contents to help and enrich users' daily life.
Different scenarios hold distinct and unique characteristics, calling for domain-specific investigations and corresponding designed recommender systems.
Therefore, in this paper, we focus on food delivery recommendations to unveil unique features in this domain, 
where users order food online and enjoy their meals shortly after delivery. 
We first conduct an in-depth analysis on food delivery datasets. 
The analysis shows that
repeat orders are prevalent for both users and stores, and situations' differently influence repeat and exploration consumption in the food delivery recommender systems. 
Moreover, we revisit the ability of existing situation-aware methods for repeat and exploration recommendations respectively, and find them unable to effectively solve both tasks simultaneously.
Based on the analysis and experiments, we have designed two separate recommendation models -- \textit{ReRec} for repeat orders and \textit{ExpRec} for exploration orders; both are simple in their design and computation.
We conduct experiments on three real-world food delivery datasets, and our proposed models outperform various types of baselines on repeat, exploration, and combined recommendation tasks.
This paper emphasizes the importance of dedicated analyses and methods for domain-specific characteristics for the recommender system studies.
\end{abstract}



\maketitle

\section{Introduction}
\label{sec:intro}

Recommenders have become pervasive in our daily life, ranging from e-commerce and advertising to music and social media.
Extensive research has been conducted on scenario-agnostic generic recommender systems, providing valuable insights for understanding user preferences in common.
However, significant variations exist in task settings and user behavior patterns across different recommendation scenarios. 
For instance, delivery location is a required input for food delivery recommenders, while location is not essential for music recommendation.
Regarding user behaviors, repeat consumption of the same items is common in grocery shopping~\cite{ariannezhad2022recanet}, while users usually do not read similar news repeatedly. 
Analyzing scenario-specific characteristics and designing corresponding recommenders is valuable for both research and practice.

In this paper, we take the food delivery scenario as a case study
to investigate user behavior patterns and influential factors.
Based on the investigation, dedicated recommenders are further designed to make better use of the scenario-specific characteristics.

In this specific and important scenario, where users order food online and enjoy their meals after delivery, is selected for our case study for two reasons.
First, food delivery platforms have gained widespread popularity globally, with hundreds of Apps providing services for billions of users.\footnote{\url{https://www.businessofapps.com/data/food-delivery-app-market/}} It is one of the recommendation services that has significantly benefited our daily lives. 
Second, food delivery is an interesting research scenario: Recommendations are naturally situation-sensitive since users' food preferences usually vary with time; and delivery distance is also an important factor for when the food can arrive. 
Moreover, people often make similar food choices, which may contribute to noticeable repeat consumption.
These characteristics distinguish food delivery recommendations from many others, making it worthy of in-depth study.

However, speculations about ubiquitous repeat and exploration consumption, as well as the impact of situations for repeat and new consumption, have not been confirmed with real data on food delivery platforms.
In this paper, we conduct in-depth analyses of food delivery datasets for three cities located in Asia and Europe from two food delivery platforms, as shown in Section~\ref{sec:dataAnalysis}. Our analyses show similar findings in interactions across all three datasets, including: 
(i) Repeat and exploration orders are both prevalent for users and stores.\footnote{As explained in Section~\ref{sec:problem}, the recommended items in this study refer to stores (\eg restaurants) rather than specific food orders (\eg the dishes ordered from a restaurant).} 
(ii) Historical situations exert a stronger influence on users' choices for repeat consumption compared to exploration consumption. (iii) Collaborative situations, on the contrary, have a greater impact on users' exploration consumption.

Building upon these findings, we re-evaluate the efficacy of current situation-aware recommenders in recommending repeat and new stores in the food delivery scenario in Section~\ref{ssec:revisitSolutions}. Results show that sequential recommenders shine for repeat consumption, while collaborative filtering-based recommenders achieve better recommendations for exploration, consistent with our analyses.
More importantly, recommending new stores for users is considerably more challenging than suggesting repeat stores. 
Thus, integrating the modeling or evaluation of repeat and exploration consumption may result in overlooking the exploration recommendation task. 
In fact, we argue that the two tasks are fundamentally different due to distinct candidate items, \ie the stores a user has interacted with before (a small set) the versus stores a user has never interacted with (a much larger set). 
Nevertheless, existing solutions, designed for generic recommendation tasks, do not adequately address such specific factors (\ie importance of situations and distinct repeat/exploration consumption) crucial for food delivery recommendations.

Inspired by the above findings, we propose two simple yet effective situation-aware methods for repeat and exploration recommendations, respectively, in Section~\ref{sec:model}.
The repeat recommender, \textbf{RepRec}, relies on the similarity between historical and current situations to recommend stores from historical interactions. 
The exploration recommender, \textbf{ExpRec}, utilizes various information, especially global situation representation and collaborative users, as exploration triggers to recommend new items.
In practice, the repeat and exploration recommendations can be presented to users distinctly, such as through segregated display cards or slots. 
This proactive approach not only ensures a streamlined user experience but also acknowledges the unique nature of each recommendation type. 
Whereas, if a unified recommendation list is needed for App displa, an intent-aware ensemble module is proposed to aggregate the repeat and exploration recommendations. 
Extensive experiments have been conducted on real-world food delivery datasets from three Asian and European cities released by two platforms.
Our proposed simple and efficient models, RepRec, ExpRec, and Ensemble module, outperform the baselines of various types in respective tasks. 

The main contributions of this paper are summarized as follows.
\begin{itemize}[nolistsep, leftmargin=*]
\item We conduct \textit{an in-depth analysis on real-world food delivery datasets} covering three cities in different countries, which reveals ubiquity of repeat orders, distinct influences of situations on repeat and exploration consumption, and necessity for separate design for repeat and exploration in food delivery recommendations.
\item We formulate the food delivery recommendation task as two sub-tasks and \textit{design recommenders for repeat consumption and exploration consumption respectively} based on scenario-specific requirements.
\item We conduct experiments to \textit{evaluate our simple recommenders against state-of-the-art models} considering repeat/exploration patterns and situation-awareness. Results show that our recommenders, proposed for the practical scenario, beat generic models with better efficiency.
\end{itemize}
\vspace{-0.2cm}

\section{Related Work}
\label{sec:related}
In recommendation systems, users' intentions for repeat consumption vary across scenarios~\cite{anderson2014dynamics}. 
We review related studies on repeat and exploration in Section~\ref{ssec:relatedRvE}. Empirical studies also demonstrate that users' preferences are influenced by dynamic situations, including factors such as location, temporal information, and activities~\cite{li2022towards,lin2023exploring}. We review the related studies in Section~\ref{ssec:relatedSR}. 

\subsection{Repeat and Exploration in Recommendation}
\label{ssec:relatedRvE}

Repeat-aware recommenders have been widely discussed in  music listening~\cite{tsukuda2020explainable,reiter2021predicting}, e-commerce~\cite{wang2019modeling,quan2023enhancing}, and grocery shopping~\cite{ariannezhad2022recanet,katz2022learning} scenarios.
Some studies focus on modeling the repeat patterns of users or items in historical sequences of interactions. Examples include predicting the interest sustainability of items~\cite{hyun2020interest}, modeling the item necessity intensity with random point process~\cite{wang2019modeling}, and enhancing item embedding with corresponding personalized or global repeat-consumption intervals~\cite{ariannezhad2022recanet,rappaz2021recommendation,quan2023enhancing}.
Another line of research pays attention to separating repeat consumption recommendation from all by specific modules.
For instance, in RepeatNet~\cite{ren2019repeatnet}, a repeat mode and an explore mode are respectively proposed to decode users' repeat and exploration preferences from a shared user sequential embedding.
NovelNet~\cite{li2022modeling} utilizes a pointer network to model repeat consumption behaviors and an attention module for exploration recommendation.

When it comes to recommending new items to users, diverse perspectives have emerged.
One is serendipity recommendation, which explore items that are unexpected yet match the users' interests~\cite{kotkov2023rethinking}.
A key focus of these recommenders lies in modeling the possible directions of new interest for users.
For example, serendipity vectors are generated to represent users' exploration preferences in \cite{li2020directional}.
\citet{zhang2021snpr} use the similarity of time and locations to provide serendipity locations for the next POI recommendation.
Review text is also adopted to measure and predict serendipity of items~\cite{fu2023wisdom}.
Another view on exploration in recommendations comes from Reinforcement learning~(RL) methods~\cite{afsar2022reinforcement}, where the dynamic exploration process of users are simulated with an RL environment. 
However, the majority of RL-based recommenders utilize interaction-level training and test settings, which are entirely different from mainstream recommendation models.

Analyses in sequential~\cite{li2023next} and next-basket recommendation~\cite{li2022repetition} show a trade-off between repeat and exploration in recommendations.
Previous studies have predominantly focused on either exploration or repeat, often lacking an analysis of the similarities and differences between these two types of interactions. Furthermore, these studies did not consider the impact of situation/context, especially dynamic context, on repeat and exploration behaviors—crucial factors in food delivery systems. In this paper, we conduct an in-depth analysis of the distinct roles of situations in repeat and exploration, and design corresponding methods for both tasks.

\subsection{Situation-aware Recommendation}
\label{ssec:relatedSR}


Context-Aware Recommender Systems~(CARS) can be applied for modeling situations in user interactions.
CARS can be categorized into two types based on the way of information utilization: CF-enhanced models and sequential-based models.
CF-enhanced CARS models the relations between features, such as second-order interaction terms in FM~\cite{rendle2010factorization}, attention-based pooling in AFM~\cite{xiao2017attentional}, and two-stream MLP in FinalMLP~\cite{mao2023finalmlp}. 
Sequential CARS leverages users' historical behaviors and context to predict users' preference~\cite{zhou2018deep}, preference evolving~\cite{zhou2019deep}, and long-short term interest~\cite{cao2022sampling}. 
Recent studies have paid attention to some specific types of situations.
For example, temporal features are widely considered in sequential recommenders, such as the time intervals of interactions~\cite{wang2019modeling}, multi-scale temporal embeddings~\cite{cho2020meantime}, and dynamic user profiles with semantic time features~\cite{rashed2022context}. 
Spatial situations are also employed in POI recommendation and location-based services, where they capture dynamic changes in preferences~\cite{lan2023spatio} or model complex relationships between time, locations, and users~\cite{lin2023exploring,lai2023multi}.

A series of recent works paid attention to recommenders in the food delivery scenarios.
The main focus is to employ spatial-temporal features.
For example, food delivery duration is predicted and used to enhance recommendations in~\cite{wang2021fulfillment}.
Various popular neural network architectures are integrated with time-spatial aware structures to capture the situation-sensitive preference changes, including multi-head attention~\cite{lin2023exploring}, multi-layer transformation~\cite{du2023basm}, contrastive learning~\cite{jiang2023cspm}, and graph-based embedding~\cite{zhang2023modeling}.

Existing situation-aware recommenders often lack thorough analysis and modeling of user behavior patterns, particularly in distinguishing between repeat and new consumption. The distinctions and similarities in the impact of situations on repeat and exploration consumption remain unclear. Recognizing the significance of both repeat consumption patterns and situations in food delivery recommendations, we conduct a comprehensive analysis that considers both aspects. Based on this analysis, we propose simple yet effective models to enhance performance for both repeat and exploration recommendation tasks.

\section{The Research Problem}
\label{sec:problem}


In food delivery services, an interaction is recorded when a user initiates an order for food from a store, providing details such as the desired food, time, and location of delivery. Formally, 
\begin{definition}[Food Delivery Interaction]\label{def:interaction}
A food delivery interaction $x$ is a 4-tuple $x=(u,s,t,\ell)$, representing user $u$ ordered food from store $s$ at time $t$ with delivery location $\ell$. We denote the set of all interactions as $\mathcal{X}$. 
\end{definition}
Note that location $\ell$ refers to the location where the food shall be delivered to. This location is usually different from the store's physical location, which is an attribute of the store.  

Because store recommendation is considered an urgent practical task in food delivery applications~\cite{zhang2023modeling}, we focus on recommending \textit{stores} instead of food to users, \ie stores rather than food are treated as items in recommender systems. 
Accordingly, we define the research problem as follows. 
\begin{definition}[Food Delivery Recommendation]\label{def:rec}
Given a set of historical food delivery interactions $\mathcal{X}$, the task is to recommend a list of stores $\{s_i\}_{i=1}^N \subset S$ for a user $u$ when she needs food recommendations at time $t$ with delivery location $\ell$.
\end{definition}
In our problem definition, in addition to user id $u$, both time $t$ and delivery location $\ell$ are known to the recommendation model as input, which are denoted as situations. Such setting simulates the practical scenario of food delivery service: the recommender works only when a user $u$ opens a food delivery app at time $t$ and specifies delivery location $\ell$.

\section{Repeat, Exploration, and Situations} 
\label{sec:dataAnalysis}

To analyze common behavior patterns and preferences among users in the food delivery scenario, we leverage datasets collected by two renowned platforms across three distinct cities worldwide: 
    
\paratitle{Delivery Hero Recommendation Dataset (\textbf{DHRD})}~\cite{assylbekov2023delivery}. The DHRD dataset comprises data from over 4.5 million orders placed by more than 1 million users on the Delivery Hero platform spanning a period of three months.
    We utilize the data of two cities in our research, namely \textbf{DHRD-SE} of Stockholm, and \textbf{DHRD-SG} of Singapore, respectively.\footnote{Data of Taipei was not used in this research due to a timezone mismatch in the data.} We also refer to the datasets as \textbf{D-SE} and \textbf{D-SG}, respectively, for short, when the context is clear.
    
    \paratitle{Takeout Recommendation Dataset (\textbf{TRD})}.\footnote{https://zenodo.org/records/8025855} The TRD dataset encompasses one-month orders from 0.2 million users in Beijing using the Meituan Takeout app.

To observe stable user behavior patterns in  the datasets, we filter users with fewer than ten orders in DHRD, and fewer than five orders in TRD, based on the data distribution. 
Statistics of the filtered datasets used in our experiments are reported in Table~\ref{tab:dataset_statistics}.

\begin{table}
    \centering
    \setlength{\abovecaptionskip}{0.cm}
    \setlength{\belowcaptionskip}{0.cm}
    \caption{Statistics of three datasets from three cities around the world on two platforms. Rep.\% represents the percentage of repeat consumption. }
    \label{tab:dataset_statistics}
    \begin{tabular}{l|rrrr|l}
    \toprule
    Dataset & \#User    & \#Store & \#Order &Rep.\% & City \\
    \midrule
    DHRD-SE  & 7,900 & 1,131 & 128,329&  56.5\% &Stockholm \\
    DHRD-SG  & 46,129 & 7,296 & 791,088& 42.3\% & Singapore\\
    TRD (Meituan) &  76,118 & 3,169 & 839,011 & 38.0\% &Beijing \\
    \bottomrule
    \end{tabular}
    \vspace{-0.2cm}
\end{table}

\subsection{Noticeable Repeat Consumption}
\label{ssec:obvRepeat}
\begin{figure*}
    \centering
    \setlength{\abovecaptionskip}{0.1cm}
    \setlength{\belowcaptionskip}{0.cm}
    \captionsetup{font=small}
    \begin{subfigure}{0.23\textwidth}
    \includegraphics[width=\linewidth]{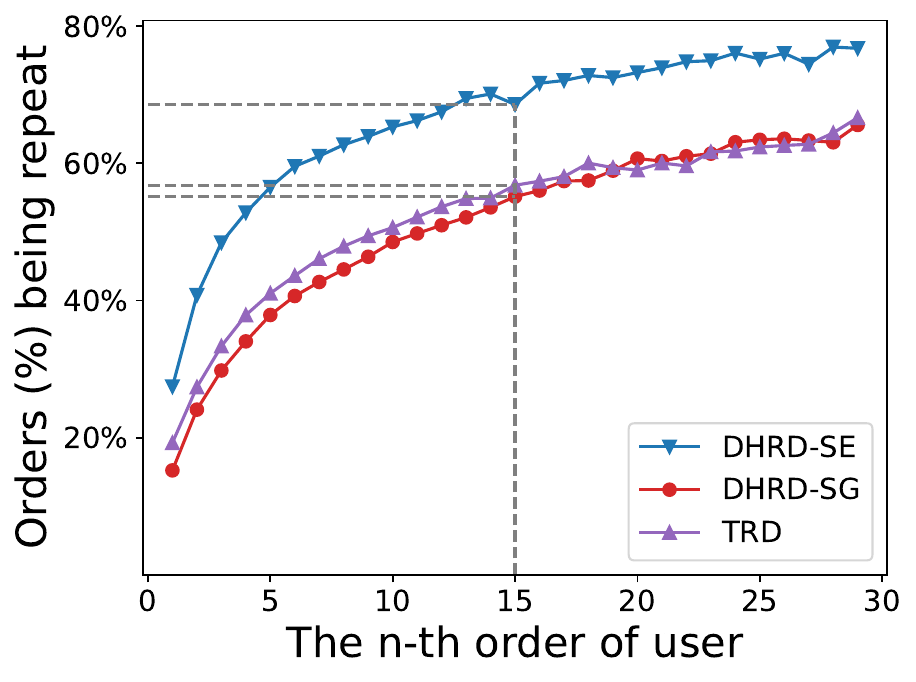} 
    \caption{Repeat ratio against orders.}
    \label{fig:repeat_a}
    \end{subfigure}
    \begin{subfigure}{0.23\textwidth}
    \includegraphics[width=\linewidth]{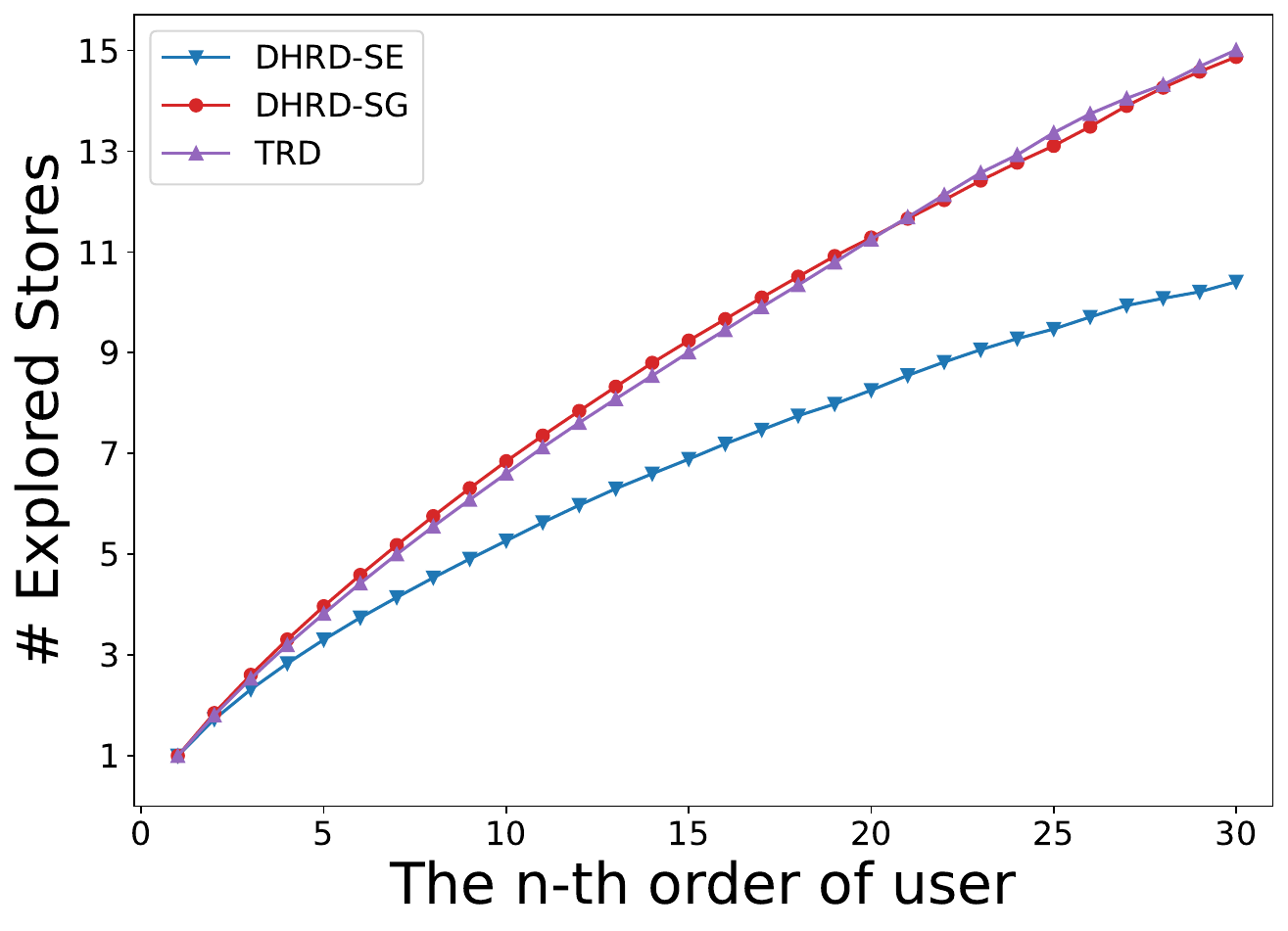} 
    \caption{Explored stores against orders.}
    \label{fig:exploration}
    \end{subfigure}
    \begin{subfigure}{0.23\textwidth}
    \includegraphics[width=\linewidth]{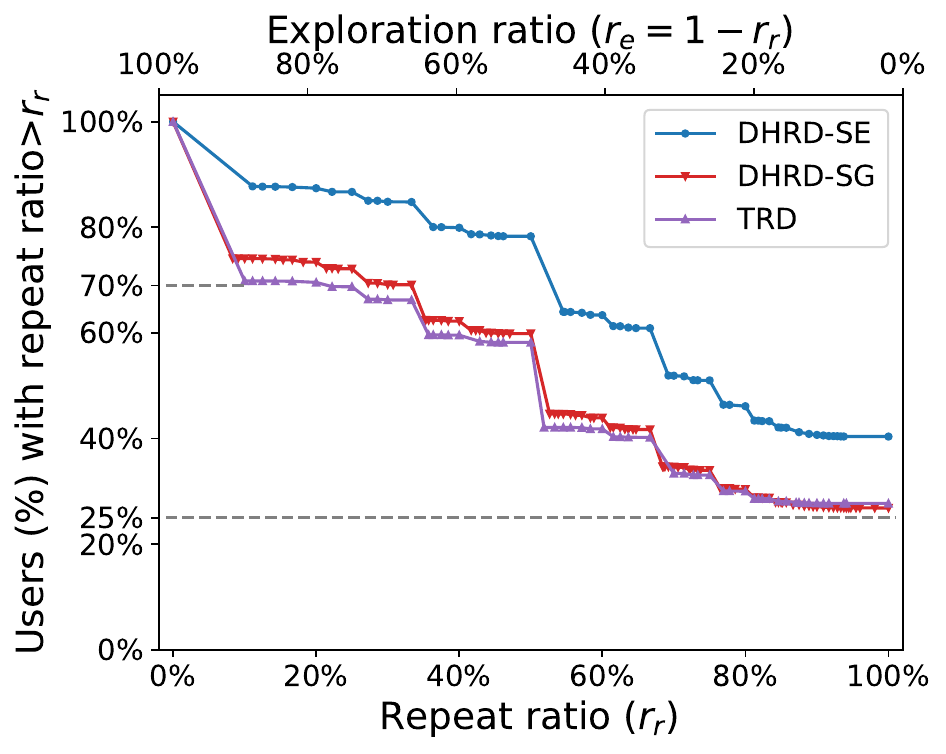} 
    \caption{Repeat/Exploration ratio of users.}
    \label{fig:repeat_b}
    \end{subfigure}
    \begin{subfigure}{0.23\textwidth}
    \includegraphics[width=\linewidth]{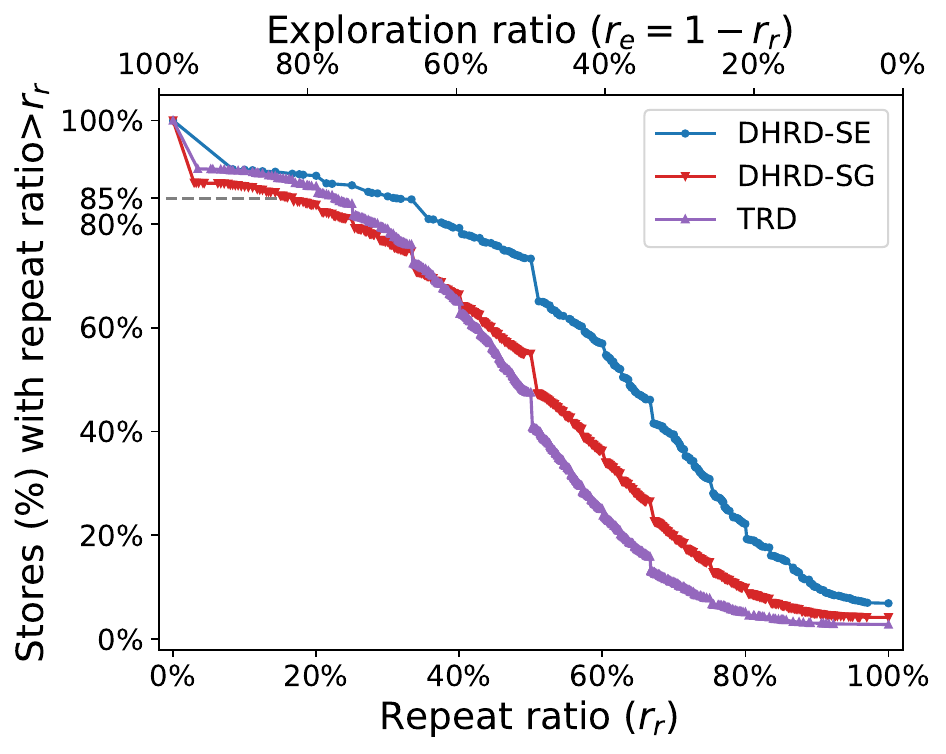} 
    \caption{Repeat/Exploration ratio of stores.}
    \label{fig:repeat_c}
    \end{subfigure}
    \captionsetup{font=normal}
    \caption{Statistics on repeat consumption: (a) The percentages of orders being repeat at the $n$-th order of all users, (b) The number of explored stores at the $n$-th order of all users, (c) The cumulative distribution of repeat/exploration ratio in the last two weeks of users, and (d) that of all stores. }
    \label{fig:repeat}
    \vspace{-0.2cm}
\end{figure*}

People are creatures of habit, and having a habitual diet is common in our daily lives.
Therefore, in the food delivery scenario, a reasonable conjecture is that \textit{users often repeatedly place orders from the same store}, \ie the existence of \textbf{repeat consumption}. As our key focus is store recommendation, we generalize the concept of ``same food'' to ``same store''.
And all orders except for repeat ones are defined as \textbf{exploration consumption}.

To validate our conjecture on repeat consumption, we plot the percentage of repeat consumption of users' $n$-th order~($1\leq n\leq 30$) in Figure~\ref{fig:repeat_a}. 
At fifteenth order ($n=15$), 68.6\% Stockholm users, 55.1\% Singapore users, and 56.8\% Beijing users choose a store which they have previously ordered from (\ie from the first 14 orders). 
We also show the number of stores that users have already explored in Figure~\ref{fig:exploration}, revealing that while the number of explored stores grows with orders, it remains limited compared to the entire dataset.
We further calculate the ratio of repeat and exploration consumption~(denoted as $r_r$ and $r_e$, respectively) for orders placed in the last two weeks. 
The \textit{cumulative distributions} of users and stores with repeat~(exploration) ratio greater~(less) than $r_r$~($r_e$) are shown in Figures~\ref{fig:repeat_b} and~\ref{fig:repeat_c}, respectively.
It shows that over 70\% users and over 85\% stores have experienced repeat consumption in the last two weeks.
On the other, almost all stores had new customers, and more than half of users explored new stores. 
Nevertheless, compared with general sequential recommendation datasets~\cite{li2022repetition}, the repeat ratio in the food delivery scenario is noticeable.

\begin{observation}\label{obs:repeat}
Repeat and exploration orders are both prevalent in the food delivery scenario among users and stores.
\end{observation}

Observation~\ref{obs:repeat} emphasizes users' noticeable repetition habits in food delivery.
Moreover, as the explored stores are limited in users' orders and users' unexplored stores are much more than consumed ones, the candidate search space for repeat recommendation is much smaller than exploration recommendation. 
Therefore, recommending for repeat and exploration should be treated separately in the food delivery scenario.

However, there is a lack of in-depth analysis of distinct characteristics of users' preferences for repeat and exploration consumption, as well as how to design corresponding recommendation strategies.
As users' situations~(time and location) are important factors for food selections, we explore the disparities in situations' influence on user choices between repeat and new consumption. 
Next, we examine the influences of situations from two common aspects in recommendation: users' own history and collaborative information from similar users, in Sections~\ref{ssec:obvHistSituations} and~\ref{ssec:obvCollaSituations}, respectively.
\vspace{-0.2cm}

\subsection{Historical Situations}
\label{ssec:obvHistSituations}

\begin{figure*}
    \centering
    \setlength{\abovecaptionskip}{0cm}
    \setlength{\belowcaptionskip}{0cm}
    \begin{subfigure}{0.31\textwidth}
    \includegraphics[width=\linewidth]{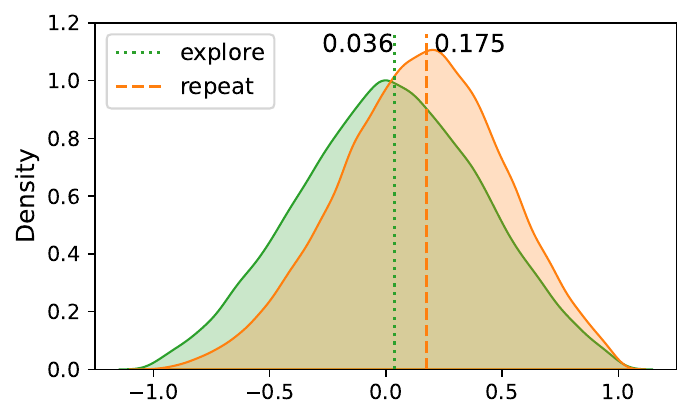} 
    \caption{DHRD-SE (Stockholm)}
    \label{fig:historical_situ_se}
    \end{subfigure}
    \begin{subfigure}{0.31\textwidth}
    \includegraphics[width=\linewidth]{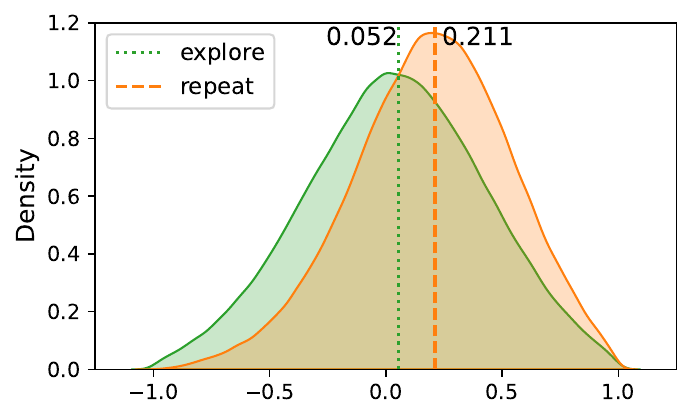} 
    \caption{DHRD-SG (Singapore)}
    \label{fig:historical_situ_sg}
    \end{subfigure}
    \begin{subfigure}{0.31\textwidth}
    \includegraphics[width=\linewidth]{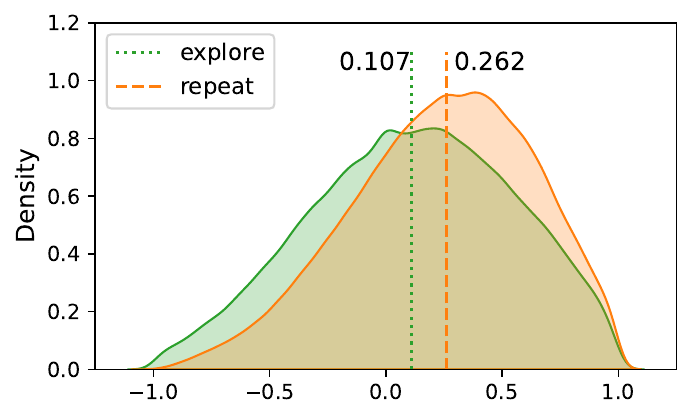} 
    \caption{TRD (Beijing)}
    \label{fig:historical_situ_mt}
    \end{subfigure}
    \caption{Distributions of the influences of historical situations, $Inf_{his}$, on repeat and exploration consumption, where influences are quantified with the correlations between historical situation similarities and store similarities. 
    }
    \label{fig:historical_situ_all}
\end{figure*}

Users' own interaction history reflects their preferences and habits, which inspire various sequential recommendation methods. 
We begin by analyzing the influences of historical situations, specifically examining the associations between users' historical situations and store selections for repeat and exploration consumption.

We quantify the association between historical situations and stores through the correlations between historical situation similarity and store similarity. 
Given an interaction $x=(u,s,t,\ell)$ of user $u$ at time $t$, let all historical interactions of user $u$ be $\vec{x}_u=\{x_1,...,x_n\}$ with $x_i=(u,s_i,t_i,\ell_i), t_i<t$.
The situation similarity between current $x$ and a historical interaction $x_i\in\vec{x}_u$ is measured by the differences in time and locations.
For time difference, the date, hour, and day of week are extracted from the timestamps $t$ and $t_i$, and the differences are calculated and normalized for each aspect.
As for locations, since the datasets only present encoded IDs, we assess whether location IDs are the same for $x$ and $x_i$.
\begin{equation}
    Sim_{\mu,i}=1-\frac{1}{4}\left(D(date)+D(hour)+D(dayOfWeek)+\mathbf{I( \ell = \ell_i)}\right)
    \label{eq:simSituation}
\end{equation}
Here, $D$ indicates the normalized difference, and $\mathbf{I}$ is the indicator function.
Store similarity of $s$ and $s_i$ is calculated by the differences of store attributes, including the store brand/chain, primary cuisine types, and location IDs.
\begin{equation}
    Sim_{\tau,i} = \frac{1}{3}\sum_{k=1}^3\mathbf{I}(s^k=s_i^k)
    \label{eq:simStore}
\end{equation}
Here, $s^k$ denotes the $k$-th attribute of store $s$.
Finally, the influence of historical situations, $Inf_{his}$, is indicated by Pearson correlation between the situation similarity sequence $\{Sim_{\mu,i}\}_{i=1}^n$ and the store similarity sequence $\{Sim_{\tau,i}\}_{i=1}^n$.
Higher $Inf_{his}$ indicates closer relations between situation similarity and store preference similarity, \ie greater influence of historical situations on store selections.

The distribution densities of $Inf_{his}$ of all repeat and exploration consumption in three datasets are shown in Figure~\ref{fig:historical_situ_all}.\footnote{Consumption with at least five historical interactions is included to avoid unstable correlations.}
Despite the diverse culinary cultures and different food delivery platforms in the three cities corresponding to datasets, the distinctions between repeat and exploration consumption remain consistent:
The influences of historical situations are notably higher for repeat consumption than exploration on average. Moreover, repeat interactions are less likely to exhibit negative correlations between historical situations and stores. In other words, for repeat consumption, users tend to order food from similar stores when they find themselves in similar situations as in the past. However, the influence of historical habits tends to weaken when exploring new stores.

\vspace{-0.1cm}
\begin{observation}\label{obs:historicalSituation}
In general, historical situations exert a greater influence on users' choices for repeat consumption compared to exploration consumption.
\end{observation}

\begin{figure*}[h]
    \centering
    \setlength{\abovecaptionskip}{0cm}
    \setlength{\belowcaptionskip}{0.cm}
    \begin{subfigure}{0.31\textwidth}
    \includegraphics[width=\linewidth]{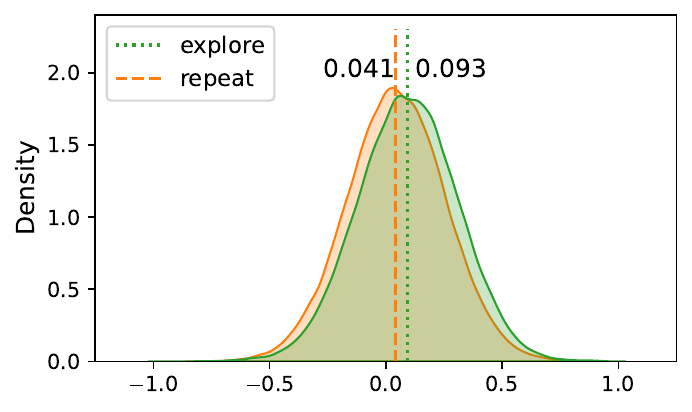} 
    \caption{DHRD-SE (Stockholm)}
    \label{fig:neighbor_situ_se}
    \end{subfigure}
    \begin{subfigure}{0.31\textwidth}
    \includegraphics[width=\linewidth]{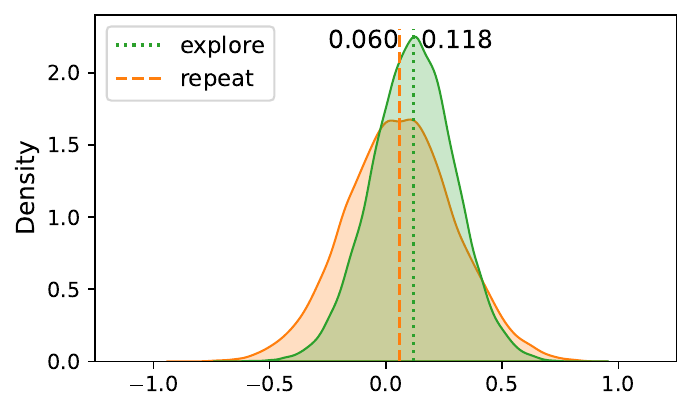} 
    \caption{DHRD-SG (Singapore)}
    \label{fig:neighbor_situ_sg}
    \end{subfigure}
    \begin{subfigure}{0.31\textwidth}
    \includegraphics[width=\linewidth]{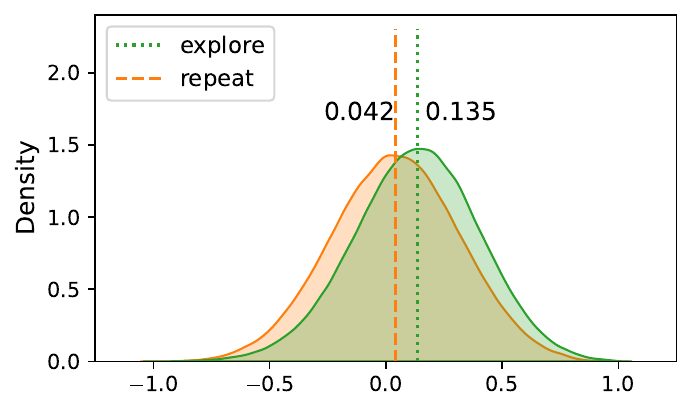} 
    \caption{TRD (Beijing)}
    \label{fig:neighbor_situ_mt}
    \end{subfigure}
    \caption{Distributions of the influences of collaborative situations, $Inf_{col}$, on repeat and exploration consumption. Influences are quantified with the correlations between situation similarities and store similarities of interactions of collaborative users. 
    }
    \label{fig:neighbor_situ}
    \vspace{-0.2cm}
\end{figure*}

\subsection{Collaborative Situations}
\label{ssec:obvCollaSituations}

In addition to users' own historical interactions, collaborative information from other users is generally essential for modeling user preferences in recommender systems.
We now explore the influences of collaborative situations, specifically examining the relationships between collaborative users' situations and store selections for both repeat and exploration consumption.

The \textit{collaborative users} are derived by the similarity of their store choices. Specifically, for each user, we represent her store preference using the relative frequency of each store she has interacted with before. User similarity is then calculated based on the correlation of the store preference vectors. The $K$ most similar users ($K=10$ in our analysis) are considered to be a user's collaborative users.

The influences of collaborative situations are quantified by correlations between the similarity of situations and the similarity of stores among collaborative users. 
Given a user's interaction $x=(u,s,t,\ell)$, the collaborative interaction set $X_c=\{(u_i,s_i,t_i,\ell_i)\}$ includes recent interactions made by $K$ collaborative users of $u$, with $t_i\in (t-t_\delta, t)$. In our analysis, we set $t_\delta$ to be one week.\footnote{We only consider recent interactions from all collaborative users to factor in the temporal dynamics of the user-item interactions along global timeline~\cite{SunFreshLook23}.}
Afterward, similarities of situations and stores are measured in the same way as in Eq.~\ref{eq:simSituation} and~\ref{eq:simStore}, and their Pearson correlation represents the influences of collaborative situations, $Inf_{col}$.

The distribution densities of $Inf_{col}$ of repeat and exploration consumption are shown in Figure~\ref{fig:neighbor_situ}.
Contrary to $Inf_{his}$, the influence of collaborative situations ($Inf_{col}$) on exploration consumption consistently surpasses that on repeat consumption across different datasets, indicating a greater impact of collaborative influences in exploration consumption.
In other words, interactions by collaborative users in similar situations contribute more to predicting a user's exploration to some extent.

\begin{observation}\label{obs:collaborativeSituation}
In general, collaborative situations exert a greater influence on users' choices for exploration consumption compared to repeat consumption.
\end{observation}

\subsection{Revisit Existing Solutions}
\label{ssec:revisitSolutions}

\begin{table}[]
\setlength{\abovecaptionskip}{0cm}
\setlength{\belowcaptionskip}{0cm}
\caption{Performance of CF-enhanced~(SOnly and FM) and sequential~(HisPop and DIN) recommenders, considering situations. \textit{Rep.} and \textit{Exp.} are short for Repeat and Exploration.
The best results are in boldface and second best underlined. 
}
\label{tab:revisitSolutions}
\begin{tabular}{l|cc|cccc}
\toprule
- & \multicolumn{2}{c|}{\textbf{Group/Dataset}} & \multicolumn{1}{c}{\textbf{SOnly}} & \multicolumn{1}{c}{\textbf{FM}} & \multicolumn{1}{c}{\textbf{HisPop}} & \multicolumn{1}{c}{\textbf{DIN}} \\
\midrule
\parbox[t]{2mm}{\multirow{6}{*}{\rotatebox[origin=c]{90}{HR@3}}}
& \multirow{3}{*}{Rep.} & D-SE & -- & 0.6996 & \underline{0.7059} & \textbf{0.7629} \\
 &  & D-SG & -- & \underline{0.6038} & 0.5941 & \textbf{0.6819} \\
 &  & TRD & -- & 0.5583 & \underline{0.5752} & \textbf{0.7211} \\
 \cmidrule(lr){2-7}
 &\multirow{3}{*}{Exp.}  & D-SE & \textbf{0.1010} & \underline{0.0764} & -- & 0.0639 \\
 &  & D-SG & \underline{0.1458} & \textbf{0.1888} & -- & 0.1409 \\
 &  & TRD & \textbf{0.1090} & \underline{0.0962} & -- & 0.0877 \\
 \midrule
 \midrule
\parbox[t]{2mm}{\multirow{6}{*}{\rotatebox[origin=c]{90}{NDCG@3}}}& \multirow{3}{*}{Rep.} & D-SE & -- & 0.5620 & \textbf{0.6433} & \underline{0.6229} \\
 &  & D-SG & -- & 0.4920 & \underline{0.4999} & \textbf{0.5626} \\
 & & TRD & - & 0.4619 & \underline{0.4829} & \textbf{0.5954} \\
 \cmidrule(lr){2-7}
 & \multirow{3}{*}{Exp.} & D-SE & \textbf{0.0826} & \underline{0.0533} & -- & 0.0425 \\
 &  & D-SG & \underline{0.1056} & \textbf{0.1362} & -- &  0.0996 \\
 &  & TRD & \textbf{0.0801} & \underline{0.0715} & -- & 0.0539 \\
 \bottomrule
\end{tabular}
 \vspace{-0.5cm}
\end{table}

Observations~\ref{obs:historicalSituation} and~\ref{obs:collaborativeSituation} indicate distinct influences of situations on users' repeat and exploration consumption in food delivery scenario.
Next, we explore the extent to which these distinct influences are reflected in existing situation-aware recommenders.

Situations can be viewed as context information; hence, we evaluate the performance of two classical Context-Aware Recommender Systems~(CARS): a CF-enhanced model, \textbf{FM}~\cite{rendle2010factorization}, and a sequential CARS, \textbf{DIN}~\cite{zhou2018deep}.
We also include two straightforward methods for comparison: 
Based on situations only,
\textbf{SOnly} embeds both items and situations and uses the dot product of their embeddings to recommend items under the current situation, optimized with a BPRMF~\cite{rendle2012bpr}-similar structure with situations replacing users. 
\textbf{HisPop} utilizes the weighted frequency of each store in a user's historical interactions as the probability for recommending the store. The weights are generated by situation similarity as in Eq.~\ref{eq:simSituation}.
These two methods strive for simplicity by exclusively considering the impact of collaborative and historical situations, respectively. 
In SOnly, recommendations are identical within the same situation. Conversely, HisPop relies solely on users' historical interactions, limiting recommendations to previously interacted stores.\footnote{Results of more existing recommenders are shown in Table~\ref{tab:overallPerformanceNew}, indicating similar inspiration.}

We conduct top-$k$ food delivery recommendation tasks for repeat and exploration consumption interactions separately.\footnote{
For the sake of coherence, we focus only on results and inspiration here.
We follow Section~\ref{ssec:expSettings} for dataset split, negative sampling, and model training strategy.} 
The performances of the four methods are reported in Table~\ref{tab:revisitSolutions}.

For repeat recommendation, sequential models outperform CF-based models on all three datasets. 
Notably, even HisPop, which solely relies on data statistics without training, surpasses FM in most cases.
For exploration consumption, surprisingly, the non-personalized SOnly, which represents the global preferences of the whole community, always achieve the best or second-best performances.
FM also outperforms DIN on exploration.
The results align with our earlier analysis: historical situation sequences contributes to repeat consumption more, while collaborative situations impact exploration consumption more.

Furthermore, it is not surprising that the values of metrics for repeat recommendation are significantly higher than those for exploration in Table~\ref{tab:revisitSolutions}, since repeat store recommendation involves a much smaller candidate set than exploration recommendation. 
However, existing studies on food delivery recommendations do not treat recommending repeat stores and exploration stores separately.

We emphasize that using one overall metric for a single list of recommendations, which includes a combination of repeat and exploration stores, may overlook the varying strengths of different recommenders in these two distinct tasks. 
Additionally, enhancing repeat recommendation accuracy can lead to a more substantial overall metric improvement but may potentially compromise the performance of the more challenging exploration task. 
To the best of our knowledge, though similar observations (\ie repeat consumption is easier to predict than exploration) have been made in earlier studies in sequential recommendation~\cite{li2022repetition} and next-basket recommendation~\cite{li2023next}, the recommenders remain to be designed as a unified model, rather than treating repeat and exploration as two separate subtasks. 

In summary, existing situation-aware recommenders primarily rely on either sequence-based or CF-based models, treating repeat and exploration consumption equally. These methods overlook the distinct impact of situations in repeat and exploration consumption, thus struggling to excel in both tasks simultaneously. This underscores the need to design distinct situation-aware recommenders for repeat and exploration separately to serve users better.


\section{Situation-aware Recommender}
\label{sec:model}

\begin{figure*}
    \centering
    \setlength{\abovecaptionskip}{0cm}
    \setlength{\belowcaptionskip}{0cm}
    \includegraphics[width=0.95\linewidth]{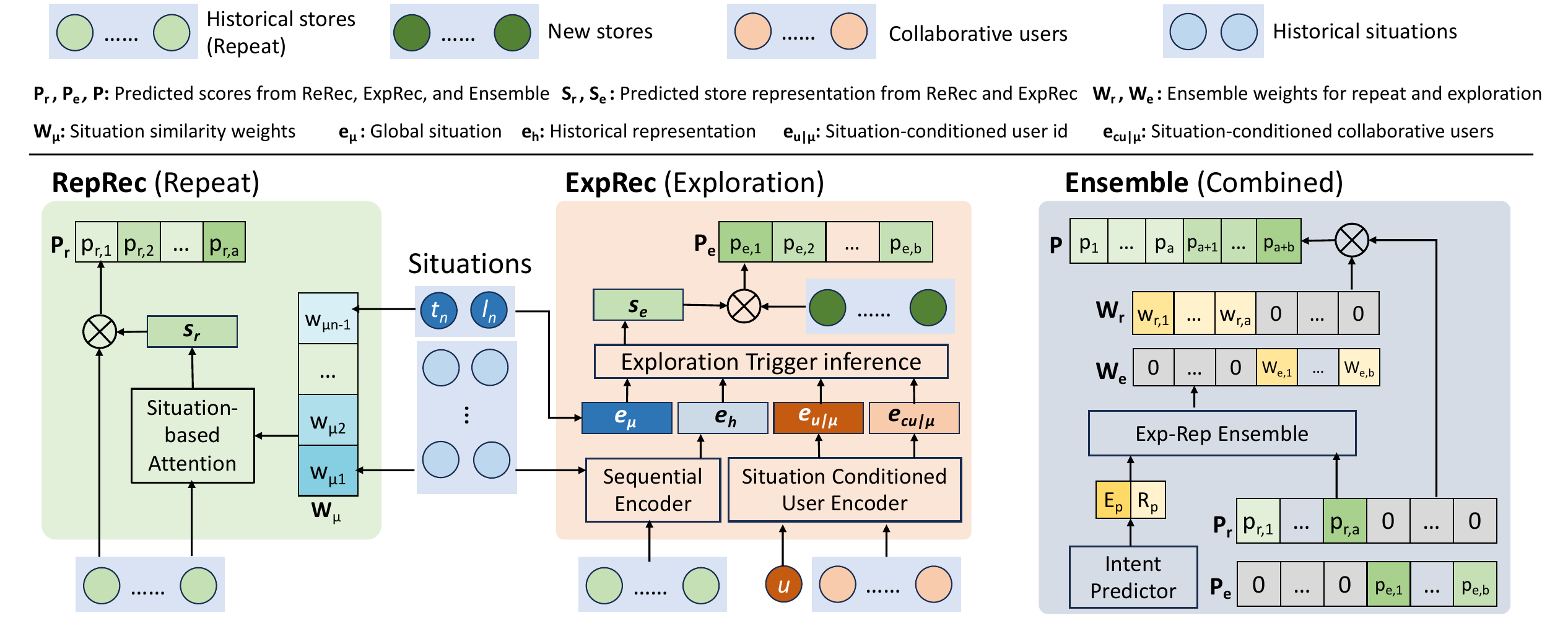}
    \caption{The framework of our proposed recommenders considering both situations and repeat/exploration patterns for food delivery. Based on the observations made in Section~\ref{sec:dataAnalysis}, \textit{RepRec} and \textit{ExpRec} are designed for repeat and exploration consumption, respectively. The \textit{Ensemble} module combines their outputs if a unified recommendation list is needed.}
    \label{fig:methodFramework}
    \vspace{-0.2cm}
\end{figure*}

Repeat consumption recommendation focuses on stores where a user has previously placed orders, typically constituting a small set.
Exploration recommendations involve suggesting new stores, and the candidate set encompasses nearly all the stores in a city (\ie except the stores a user has placed orders before). 
Given this fundamental difference in the recommendation spaces of candidate items, we approach food delivery recommendation as two distinct sub-tasks. This separation allows us to not only address the unique challenges of each sub-task but also fully leverage insights on the influences of historical situations and collaborative situations.

Accordingly, as illustrated in Figure~\ref{fig:methodFramework}, we design two situation-aware recommenders: \textbf{RepRec} for repeat consumption, and \textbf{ExpRec} for exploration consumption.
Depending on the App interface design, if only one list will be presented to users, an \textbf{Ensemble} module will combine the recommended stores from both recommenders.

\subsection{\textbf{RepRec} for Repeat Consumption}
\label{ssec:methodRepeat}

As detailed in Section~\ref{ssec:obvHistSituations}, the similarity of historical situations is correlated with the similarity of store selections for repeat consumption.
The \textbf{RepRec} for repeat consumption is designed based on this finding, illustrated on the left-hand side of Figure~\ref{fig:methodFramework}.

For a user $u$ at time $t_n$ and delivery location $\ell_n$, her historical interactions are denoted as $\{x_1,x_2,...,x_{n-1}\}$, with $x_i=(u,s_i,t_i,\ell_i), t_i<t$.
The goal of RepRec is to recommend a store from $u's$ historically interacted stores $\{s_1,s_2,...s_a\}$, $a\leq n-1$.

First, situations $(t_i,\ell_i)$ are embedded into a vector $\vec{\mu}_i$, and the attention weight of each historical interaction is illustrated with the similarity between current and historical situations in Eq.~\ref{eq:repeatSituationSimilarity}, where $\cdot$ indicates inner product, and $||$ is the norm of vector.
\begin{equation}
    \label{eq:repeatSituationSimilarity}
    w_{\mu i} = \frac{\vec{\mu}_i\cdot \vec{\mu}_n}{|\vec{\mu}_i|* |\vec{\mu}_n|}
\end{equation}

Then, an embedding layer is applied for stores, providing store representation $\vec{s}_i$.
The predicted representation for the next repeat store is generated with a situation-based attention layer,
\begin{equation}
    \label{eq:repeatAttention}
    \vec{s}_r=\sum_{i=1}^{n-1} w_{\mu i}\vec{s}_i
\end{equation}
Finally, the predicted score of each candidate store is produced by $\vec{p}_r=\{p_{r,i}=\vec{s}_i\cdot \vec{s}_r\}_{i=1}^a$.
This straightforward design increases the likelihood of recommending stores with more similar historical situations, aligning with the observed influence pattern of historical interactions on repeat consumption in Observation~\ref{obs:historicalSituation}.

\subsection{ExpRec for Exploration Consumption}
\label{ssec:methodExploration}

Exploration recommendation is a more challenging task.
Users' exploration consumption could be a result of many factors, such as promotional activities by stores, recommendations from friends, or entering a new situation, which may or may not be observable from past interactions.

As shown in the middle part of Figure~\ref{fig:methodFramework}, we consider these potential factors as triggers and incorporate them into the model input. We then utilize a trigger inference module to integrate all triggering information. 
Specifically, four types of information are considered in the exploration module: current situation $\vec{e}_\mu$, historical situations and items $\vec{e}_h$, situation-aware user representation $\vec{e}_{u|\mu}$, and situation-aware collaborative users' representation $\vec{e}_{cu|\mu}$.
Motivated by the impressive performance of SOnly in Table~\ref{tab:revisitSolutions}, the embedding of the current situation $(t_n,\ell_n)$ serves as a part of input $\vec{e}_\mu \in \mathcal{R}^{D\times 1}$.
As users' historical interactions reflect their inherent preferences, historical interactions, \ie the sequence information of historical stores and situations are concatenated and encoded with a gated recurrent unit~(GRU)~\cite{chung2014empirical} model into a vector $\vec{e}_h \in \mathcal{R}^{D\times 1}$.
For user representations, since users' preferences are influenced by situations, we propose a situation-conditioned user encoder to generate situation-aware user embedding $e_{u|\mu}$, following the light-weight conditional neural network~\cite{ramos2021conditioning}:
With $\vec{e}_\mu$ as the conditioning vector, user embedding $\vec{u}$ as the input, a pre-defined group of $M$ basic activation functions $\{A_1,A_2,...,A_M\}$~(\eg $sigmoid$, $relu$, ...), the situation-conditioned user representation is
\begin{align}
    \label{eq:expConditionUserEncoder}
    \vec{a} = softmax(\mathbf{\Theta}^\mu\vec{e}_\mu+\vec{\beta}^\mu), \quad
    \vec{e}_{u|\mu} = \sum_{j=1}^M {a}_j A_j(\vec{u})
\end{align}
Where $\mathbf{\Theta}^\mu\in \mathcal{R}^{M\times D}$ and $\vec{\beta}^\mu \in \mathcal{R}^{M\times 1}$ are learnable parameters, and ${a}_j$ denotes the $j$-th dimension of $\vec{a}$. 
For each user $u$, $K$ collaborative users are selected as in Section~\ref{ssec:obvCollaSituations}, embedded, and encoded with the situation-conditioned user encoder into $\{\vec{e}_{cu_i|\mu}\}_{k=1}^K$. Then $\vec{e}_{cu|\mu}$ is obtained by weighted summing based on similarities with $u$.

With representations of these four types of potential triggers, ExpRec aggregates them for a predicted representation $\vec{s}_e$ to recommend an exploration store with a trigger inference layer.
This layer aims to estimate the likelihood of users engaging in exploration behavior triggered by different information in the current situation.
To maintain model simplicity, 
the weights for each input trigger are directly obtained:
\begin{align}
    \vec{w}^e &= softmax\left(\mathbf{\Theta}^e(\vec{e}_\mu \oplus \vec{e}_{u|\mu})+\vec{\beta}^e\right)\\
    \vec{s}_e &= {w}^e_1 \vec{e}_\mu + {w}^e_2 \vec{e}_h + {w}^e_3 \vec{e}_{u|\mu} + {w}^e_4 \vec{e}_{cu|\mu}
\end{align}
where $\mathbf{\Theta}^e\in \mathcal{R}^{4\times D}$ and $\vec{\beta}^e\in \mathcal{R}^{4\times 1}$ are learnable parameters, and ${w}^e_i$ denotes the $i$-th dimension of $\vec{w}^e$.
Lastly, predicted exploration recommendation scores are generated by $\vec{p}_{e}=\{p_{e,i}=\vec{s}_e\cdot \vec{s}_i\}_{i=1}^b$.

\subsection{Repeat \& Exploration Combined}
\label{ssec:methodCombination}

In practice, recommendations of repeat stores and new stores can be displayed with two different cards or slots on an App, presenting diverse choices for users.
However, as most existing solutions employ a unified model to handle both repeat and exploration consumption, we devise an ensemble module to combine the recommended stores of RepRec and ExpRec for a fair comparison with existing models. 

In the combination process, determining whether a user intends to engage in repeat or exploration consumption is crucial. 
However, predicting this intent is a challenging task. Inspired by \citet{li2023intent}, in our ensemble module, we first predict users' intent, then generate item-level ensemble weights based on the predicted intent to obtain the final recommendations from prediction scores by RepRec and ExpRec.
In intent predictor, historical intents of user $u$ are encoded with a sequential encoder, concatenated with current situation embedding and user embedding, then utilized to predict the probability of repeat intent $E_p$ and exploration intent $R_p$~($E_p+R_p=1$).
Afterwards, the Exp-Rep ensemble module adopts the structure of IntEL~\cite{li2023intent}.
It applies a self-attention layer taking prediction scores $\vec{p}_e$ and $\vec{p}_r$ as input, followed by a cross-attention layer incorporating intents $(E_p, R_p)$ and a projection layer for item-level weights $\vec{w}_r$ and $\vec{w}_e$.
Finally, predictions for the $a$ repeat stores and $b$ new stores are generated,
\begin{equation}
    \label{eq:ensemblePrediction}
    p_i = \left\{
    \begin{array}{rcl}
    w_{r,i}\cdot p_{r,i} & & i\leq a\\
    w_{e,i-a}\cdot p_{e,i-a} & & i>a
    \end{array}
    \right.
\end{equation}

\section{Experiments}
\label{sec:exp}

\subsection{Experimental Setup}
\label{ssec:expSettings}

\begin{table*}[t]
\setlength{\abovecaptionskip}{0cm}
\caption{
Top-3 recommendation performances on repeat, exploration, and combined candidate sets on three datasets.
The best results are in boldface, and second best underlined. */** indicate $p\leq 0.05/0.01$ for the paired t-test of \textit{Ours} and the best baseline.
Results of RepRec, ExpRec, and the Ensemble recommendations are shown as \textit{Ours} in each test group, respectively.}
\label{tab:overallPerformanceNew}
\small
\begin{tabular}{ccc|>{\centering}m{0.06\textwidth}>{\centering}m{0.06\textwidth}>{\centering}m{0.06\textwidth}>{\centering}m{0.06\textwidth}>{\centering}m{0.06\textwidth}>{\centering}m{0.06\textwidth}>{\centering}m{0.06\textwidth}>{\centering}m{0.06\textwidth}>{\centering}m{0.08\textwidth}|l}
\toprule
\multicolumn{3}{c|}{\textbf{Setting}} & \multicolumn{3}{c}{\textbf{Repeat-aware Models}} & \textbf{Seren.} & \multicolumn{2}{c}{\textbf{CF-based CARS}} & \multicolumn{2}{c}{\textbf{Seq.-based CARS}} & \textbf{Food Delivery} & \multirow{2}{*}{\textbf{Ours}}  \\ 
\cmidrule(lr){4-6} \cmidrule(lr){7-7}\cmidrule(lr){8-9} \cmidrule(lr){10-11} \cmidrule(lr){12-12}
\multicolumn{3}{c|}{\textbf{Group/Dataset}} &\textbf{ReCANet} & \textbf{RepeatNet} & \textbf{TSRec} & \textbf{SNPR} & \textbf{FM} & \textbf{FinalMLP} & \textbf{DIN} & \textbf{SDIM} & \textbf{DPVP} &  \\
 \midrule

\parbox[t]{1mm}{\multirow{6}{*}{\rotatebox[origin=c]{90}{\underline{\textbf{Repeat}}}}} & \multirow{3}{*}{\rotatebox[origin=c]{90}{HR}} & D-SE & 0.7702 & 0.7694 & \underline{0.7753} & 0.7239 & 0.6996 & 0.7057 & 0.7629 & 0.7709 & 0.6768 & \textbf{0.7833} \\
&  & D-SG & 0.6626 & 0.6682 & 0.6483 & 0.6295 & 0.6038 & 0.6016 & \underline{0.6819} & 0.6792 & 0.6812 & \textbf{0.7092}**  \\
 &  & TRD & 0.7181 & 0.7235 & \underline{0.7469} & 0.6402 & 0.5583 & 0.5725 & 0.7211 & 0.7357 & 0.7183 & \textbf{0.7585}* \\
 \cmidrule(lr){3-13}
 & \multirow{3}{*}{\rotatebox[origin=c]{90}{NDCG}} & D-SE & 0.6389 & 0.6365 & \underline{0.6478} & 0.5968 & 0.5620 & 0.5710 & 0.6229 & 0.6268 & 0.5504 & \textbf{0.6588}*  \\
  &  & D-SG & 0.5435 & 0.5459 & 0.5015 & 0.5412 & 0.4920 & 0.4766 & \underline{0.5626} & 0.5581 & 0.5596 & \textbf{0.5932}**  \\
 &  & TRD & 0.5837 & 0.5853 & \underline{0.6079} & 0.5481 & 0.4619 & 0.4884 & 0.5954 & 0.5994 & 0.5459 & \textbf{0.6346}*  \\
 \midrule
 \midrule
\parbox[t]{1mm}{\multirow{6}{*}{\rotatebox[origin=c]{90}{\underline{\textbf{Exploration}}}}} & \multirow{3}{*}{\rotatebox[origin=c]{90}{HR}} & D-SE & - & 0.0448 & 0.0838 & \underline{0.1098} & 0.0764 & 0.0770 & 0.0639 & 0.0740 & 0.0939 & \textbf{0.1315}** \\
 &  & D-SG & - & 0.1139 & 0.1609 & 0.1890 & 0.1888 & \underline{0.2042} & 0.1409 & 0.1791 & 0.1433 & \textbf{0.2407}**  \\
 &  & TRD & - & 0.0755 & 0.0699 & \underline{0.1211}  & 0.0962 & 0.1181 & 0.0877 & 0.1140 & 0.1035 & \textbf{0.1463}**  \\
 \cmidrule(lr){3-13}
 & \multirow{3}{*}{\rotatebox[origin=c]{90}{NDCG}} &
 D-SE & - & 0.0259 & 0.0581 & \underline{0.0863} & 0.0533 & 0.0530 & 0.0425 & 0.0524 & 0.0698  & \textbf{0.0984}** \\
 &  & D-SG & - & 0.0745 & 0.1128 & 0.1442 & 0.1362 & \underline{0.1601} & 0.0996 & 0.1377 & 0.1091 &  \textbf{0.1768}** \\
 &  & TRD & - & 0.0512 & 0.0365 & \underline{0.0925} & 0.0715 & 0.0878 & 0.0539 & 0.0812 & 0.0805 & \textbf{0.1097}**  \\
 \midrule
 \midrule
\parbox[t]{1mm}{\multirow{6}{*}{\rotatebox[origin=c]{90}{\underline{\textbf{Combined}}}}} & \multirow{3}{*}{\rotatebox[origin=c]{90}{HR}} & D-SE & 0.5217 &  0.5272 & \underline{0.5257} & 0.4764 & 0.4364 & 0.4371 & 0.5149 & 0.5030 & 0.4581 & \textbf{0.5314} \\
 &  & D-SG & 0.3422 & 0.3674 & 0.3493 & 0.3402 & 0.3338 & 0.2920 & \underline{0.3900} & 0.3775 & 0.3618 & \textbf{0.4124}** \\
 &  & TRD & 0.3783 & 0.3838 & 0.3815 & 0.3463 & 0.2786 & 0.2179 & 0.3898 & \underline{0.3907} & 0.3502 &  \textbf{0.4010}* \\
 \cmidrule(lr){3-13}
 & \multirow{3}{*}{\rotatebox[origin=c]{90}{NDCG}} &
 D-SE & 0.4345 & 0.4360 & \underline{0.4530} & 0.4095 & 0.3604 & 0.3539 & 0.4322 & 0.4142 & 0.3732 &  \textbf{0.4673}* \\
 &  & D-SG & 0.2865 & 0.3001 & 0.2994 & 0.2798 & 0.2663 & 0.2289 & \underline{0.3190} & 0.2981 & 0.2829  & \textbf{0.3291}** \\
 &  & TRD & 0.3079 & 0.3103 & 0.3030 & 0.2842 & 0.2224 & 0.1688 & \underline{0.3167} & 0.3139 & 0.2754  & \textbf{0.3352}** \\
 \bottomrule
\end{tabular}
\vspace{-0.2cm}
\end{table*}

\subsubsection{Dataset Split and Evaluation Metric}
We employ the datasets introduced in Section~\ref{sec:dataAnalysis} and partition them into training, validation, and test sets following the global timeline. Interactions within the last two weeks constitute the \textbf{test set}, while those in the preceding two weeks serve as the \textbf{validation set} for DHRD-SE and DHRD-SG. The remaining interactions form the \textbf{training set}. In the TRD dataset, the last week and the preceding three days are designated as the test and validation sets, respectively.

For a fair comparison of different methods on repeat, exploration, and the combined recommendation sub-tasks, \textbf{negative samples} are separately constructed for evaluation on three sub-tasks. For repeat recommendation, the candidates are users' historically interacted stores. 
For the sub-task of exploration, 999 random new stores are sampled as negative candidates for recommendation.
In the combined recommendation sub-task, besides the current ground truth, we select users' interacted stores
and supplement them with randomly sampled new stores until the total number of candidates reaches 1000. 
In both exploration and combined recommendation sub-tasks, the recommender is expected to make the correct prediction among 1000 stores.  
We evaluate all three sub-tasks by HR@3 and NDCG@3; both are commonly used evaluation metrics.

\subsubsection{Baseline Methods}

To validate the effectiveness of our proposed methods to leverage situation information and recommend for repeat and exploration consumption, we compare with 9 baselines in various groups. Among them, ReCANet, RepeatNet, and TSRec are repeat-aware recommenders, while SNPR is a serendipity-oriented recommender. FM and FinalMLP are CF-based CARS, and DIN and SDIM are sequence-based CARS. DPVP is dedicated to food delivery recommendations.

\begin{itemize}[nolistsep, leftmargin=*]
    \item \textbf{ReCANet}~\cite{ariannezhad2022recanet} leverages interaction intervals of historical items to predict users' preferences, with a specific focus on recommending repeat items.
    \item \textbf{RepeatNet}~\cite{ren2019repeatnet} is the first neural recommender to explicitly emphasize repeat consumption, with repeat and exploration modes to separately decode from a shared historical representation.
    \item \textbf{TSRec}~\cite{quan2023enhancing} is one of SOTA repeat-aware recommenders, where both temporal pattern and sequential pattern of repeat consumption are considered.
    \item \textbf{SNPR}~\cite{zhang2021snpr} is a POI recommender to recommend serendipity locations aware of global similarities of time and spatial attributes.
  \item \textbf{FM}~\cite{rendle2010factorization} is one of the most classic CF-enhance CARS to combine attributes in first and second orders.
    \item \textbf{FinalMLP}~\cite{mao2023finalmlp} is a CF-enhanced CARS with two MLP-based structures to capture and fuse high-level feature interactions.

    \item \textbf{DIN}~\cite{zhou2018deep}, a classic sequential CARS, uses local activation units to assign weights for users' historical interactions dynamically.
    \item \textbf{SDIM}~\cite{cao2022sampling} is a SOTA sequential CARS. SDIM models users' dynamic long short-term context-aware behavior sequences with hash-based attention modules.
\item \textbf{DPVP}~\cite{zhang2023modeling} utilizes time-aware graphs and subgraphs to model users' preferences for locations, stores, and food. DPVP is a SOTA food delivery recommender with codes available.
\end{itemize}

Note that RL-based exploration recommenders are not evaluated because their interaction-based training and evaluation settings are entirely different from the rest of the models~\cite{afsar2022reinforcement}.

\subsubsection{Implementation Details}

We implement all methods with Pytorch.
Among baseline models, source codes from their original papers are utilized for RepeatNet, TSRec, and DPVP, while other methods are implemented with Rechorus~\cite{wang2020make} and FuxiCTR~\cite{zhu2021open} frameworks.
Adam is used as the optimizer, and early stop is adopted if HR@3 on the validation dataset continues to drop for 10 epochs.
We set the embedding size to 64 and training batch size to 256 for all methods.
Learning rate, weight decay, and model-related hyper-parameters are tuned for our methods and all baselines.
Codes and hyper-parameter settings are publicly available.\footnote{\url{https://github.com/JiayuLi-997/FoodDeliveryRecommender}}

\subsection{Overall Performance}

The top-3 recommendation performances of our proposed methods (named \textit{Ours} for RepRec, ExpRec, and Combined) and all baselines are reported in Table~\ref{tab:overallPerformanceNew}.
In general, our proposed methods achieve the best performances on all three sub-tasks across three datasets. The improvements over the second-best baselines are statistically significant for 16 out of 18 comparisons. 

For repeat consumption, repeat-aware models and sequence-based CARS yield comparable results. This suggests that users' repeat behavior patterns and situational context are equally important in repeat consumption recommendations. Thus, our simple RepRec, considering both factors, gains the best performance. 
Meanwhile, the only two non-sequential models, CF and FinalMLP, perform the worst. This result aligns with Observation~\ref{obs:historicalSituation} that the historical behaviors of users are essential for future repeat choices. 

As for exploration recommendation, the serendipity recommender SNPR is the best baseline for DHRD-SE and TRD, as exploration tendency is explicitly used as supervision signals. However, FinalMLP outperforms SNPR on DHRD-SG, possibly due to the crucial role of situational context in it.
Comparing CARS, the CF-based FinalMLP presents better exploration recommendations than Seq.-based SDIM, in line with Observation~\ref{obs:collaborativeSituation} that collaborative situations are important for exploration. 

At last, the combined recommendations of all candidates show different trends on different datasets. 
Although repeat-aware models and Seq.-based CARS are generally weak at exploration, they achieve the best combined results among baselines.
It is because they achieve great performances on the repeat recommendation sub-task and repeat consumption is easy and takes up a great proportion.
In all, no baseline consistently outperforms others on all three recommendation sub-tasks on any dataset.
Therefore, if evaluations are solely conducted on the combined results, the differences between repeat and exploration sub-tasks will be overlooked, and the results could be dominated by the relatively simple repeat recommendation task.
Our proposed ensemble method, based on the best repeat and exploration recommendations, achieves the best recommendation performances on combined candidates.

\subsection{Parameter Efficiency}
\label{ssec:expEfficiency}

As we claimed, RepRec and ExpRec are designed to adhere domain-specific characteristics with simple and effective structures.
To illustrate the simplicity of our methods, we compare the parameter numbers of our methods and best baselines for repeat and exploration sub-tasks in Table~\ref{tab:parameterEfficiency}.
It indicates that parameter numbers of our proposed methods, especially the RepRec, are much less than baseline models.
Therefore, in terms of efficiency, our simple methods for repeat and exploration demonstrate a high degree of parameter efficiency when compared with the best baselines. 
Therefore, our proposed methods, especially the RepRec, are parameter efficient.

\begin{table}[]
    \centering
\small
    \setlength{\abovecaptionskip}{0cm}
    \caption{Number of parameters of our methods and best baselines. (B.: Baseline, Rep.: Repeat, Exp.: Exploration.)}
    \begin{tabular}{c|cccc}
    \toprule
        \textbf{Dataset} & \textbf{RepRec} & \textbf{ExpRec} & \textbf{Best Rep. B.} & \textbf{Best Exp. B.} \\
    \midrule
         DHRD-SE & 0.14M & 1.57M & 2.03M & 4.04M \\
         DHRD-SG & 0.69M & 3.72M & 4.05M & 5.46M \\
         TRD & 0.50M & 6.90M & 9.06M & 16.56M \\
    \bottomrule
    \end{tabular}
    \vspace{-0.4cm}
    \label{tab:parameterEfficiency}
\end{table}

\subsection{Ablation Study on ExpRec}

\begin{figure}
    \centering
    \setlength{\abovecaptionskip}{0cm}
    \setlength{\belowcaptionskip}{0cm}
    \includegraphics[width=\columnwidth]{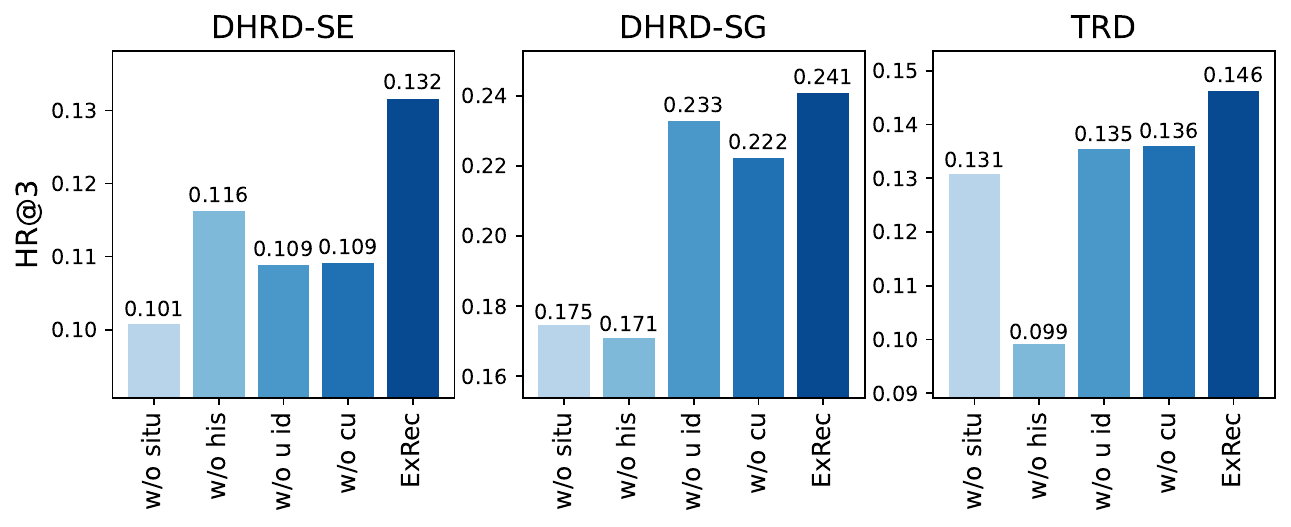}
    \caption{Performances of ExpRec and its variants without a specific trigger input. w/o is short for \textit{without}.}
    \label{fig:ablation}
    \vspace{-0.5cm}
\end{figure}

As discussed in Section~\ref{ssec:methodExploration}, various potential triggers are incorporated for exploration recommendation.
Here, we validate the importance of different triggers.
The performance on the exploration recommendation task of different variants of ExpRec, each excluding one specific trigger input at a time, is depicted in Figure~\ref{fig:ablation}.
The results indicate that removing any input leads to a decrease in performance, underscoring the importance of all triggers. 
Especially, situations emerge as crucial inputs across all three datasets, with their omission resulting in the most or second-most substantial performance decline.
Interestingly, trigger influence varies across datasets. 
User IDs and collaborative users emerge as important triggers on the DHRD-SE dataset, which has a small scale and limited stores. 
On DHRD-SG and TRD datasets with a larger candidate pool for exploration, users' historical interactions play a crucial role in revealing users' preferences.
Specifically, the user ID trigger contributes relatively less to the DHRD-SG dataset. This is because a significant portion of exploration behavior is accounted for by situational context and history, potentially diminishing the importance of user ID-related information. 

\section{Conclusion}

Behavior patterns and influencing factors vary across different recommendation scenarios. Therefore, recommender systems should be tailored to the distinctive characteristics of each specific scenario. 
In this paper, we focus on the food delivery scenario, an interesting and representative scenario where repeat consumption and situation influences are two crucial factors.
Our in-depth analysis reveals consistent observations across real-world food delivery datasets from different platforms and countries:
Repeat consumption is prevalent, where choices for repeat orders are more strongly influenced by historical situations; and exploratory selections, which are also ubiquitous, are more sensitive to collaborative situations.
Furthermore, we formally frame repeat and exploration recommendation as two distinct tasks, recognize their differences in candidate pools, and acknowledge that exploration recommendation is a much more challenging task.
Inspired by the findings, we design two simple yet effective situation-aware models, RepRec and ExpRec, respectively.
Extensive experiments indicate that our proposed methods outperform various types of baselines on repeat, exploration, and combined recommendation sub-tasks.

As emphasized, various recommendation domains display distinct task settings. Therefore, the proposed models may not be universally suitable for generic recommendation scenarios. 
However, the robust performance of our simple models compared to existing recommenders suggests a need to prioritize domain-specific characteristics in the area of recommender systems research. 
Moreover, we believe the methodology of data-driven analyses and dedicated design for recommendations will inspire more scenario-specific studies in the future.

\newpage
\bibliographystyle{ACM-Reference-Format}
\bibliography{reference}

\end{document}